\documentclass[prl,aps,reprint,twocolumn,superscriptaddress,floatfix,showpacs]{revtex4-2}

\usepackage{graphicx}
\usepackage{amssymb}
\usepackage{amsmath}
\usepackage{color}
\usepackage{changes}
\usepackage{hyperref}
\usepackage{comment}
\usepackage{lipsum}

\usepackage{ dsfont }   			    
\usepackage{float}					

\usepackage{notes2bib}
\bibnotesetup{
	note-name = ,
	use-sort-key = false
}

\newcommand{\timerev}[1]{} 


\begin{document}

\title{Quench-Probe Setup as Analyzer of Fractionalized Entanglement Spreading}
\author{Nicolas P. Bauer}
\email[Email:\,]{nicolas.bauer@physik.uni-wuerzburg.de}
\affiliation{ Institute of Theoretical Physics and Astrophysics, University of Würzburg, Germany} 
\affiliation{Würzburg-Dresden Cluster of Excellence ct.qmat, Germany}   
\author{Jan Carl Budich}
\affiliation{ Institute of Theoretical Physics, Technische Universität Dresden, Germany} 
\affiliation{Würzburg-Dresden Cluster of Excellence ct.qmat, Germany}
\author{Björn Trauzettel}
\affiliation{ Institute of Theoretical Physics and Astrophysics, University of Würzburg, Germany} 
\affiliation{Würzburg-Dresden Cluster of Excellence ct.qmat, Germany} 
\author{Alessio Calzona}
\affiliation{ Institute of Theoretical Physics and Astrophysics, University of Würzburg, Germany} 
\affiliation{Würzburg-Dresden Cluster of Excellence ct.qmat, Germany}
\date{\today}

\begin{abstract}
We propose a novel spatially inhomogeneous setup for revealing quench-induced fractionalized excitations in entanglement dynamics. In this quench-probe setting, the region undergoing a quantum quench is tunnel-coupled to a static region, the probe.
Subsequently, the time-dependent entanglement signatures of a tunable subset of excitations propagating to the probe are monitored by energy selectivity.
We exemplify the power of this generic approach by identifying a unique dynamical signature associated with the presence of an isolated Majorana zero mode in the post-quench Hamiltonian.
In this case excitations emitted from the topological part of the system give rise to a fractionalized jump of $\log(2)/2$ in the entanglement entropy of the probe.
This dynamical effect is highly sensitive to the localized nature of the Majorana zero mode, but does not require the preparation of a topological initial state.
\end{abstract}

\maketitle

\textit{Introduction.\textemdash} 
Identifying physical signatures to distinguish and understand phases of matter occurring in nature is a main objective of research in physics.
Dynamical approaches probing a system far from thermal equilibrium have become increasingly important. In particular, quantum quenches, i.e. abrupt changes of parameters in the Hamiltonian, have enabled unprecedented insights into structure and dynamics of quantum matter, both in theory \cite{Gogolin2016,Polkovnikov2011,Essler2016} and experiment \cite{Kinoshita2006,Gring2012,Cheneau2012,Langen2013,Bloch2008}.
A prominent example is provided by the prediction and observation of non-equilibrium topological invariants \cite{Hauke2014, Vajna2015, Budich2016,Goldman2016,Eckardt2017,Cooper2019,Bermudez2009,Caio2015,Hu2016,Wang2017,Sun2018,Hu2020,McGinleyCooper:,Flaschner2017,Marks2021} that probe topological properties of matter without requiring the preparation of a topological equilibrium state.

\begin{figure}[t]
	\centering
	\includegraphics[width=0.92\columnwidth]{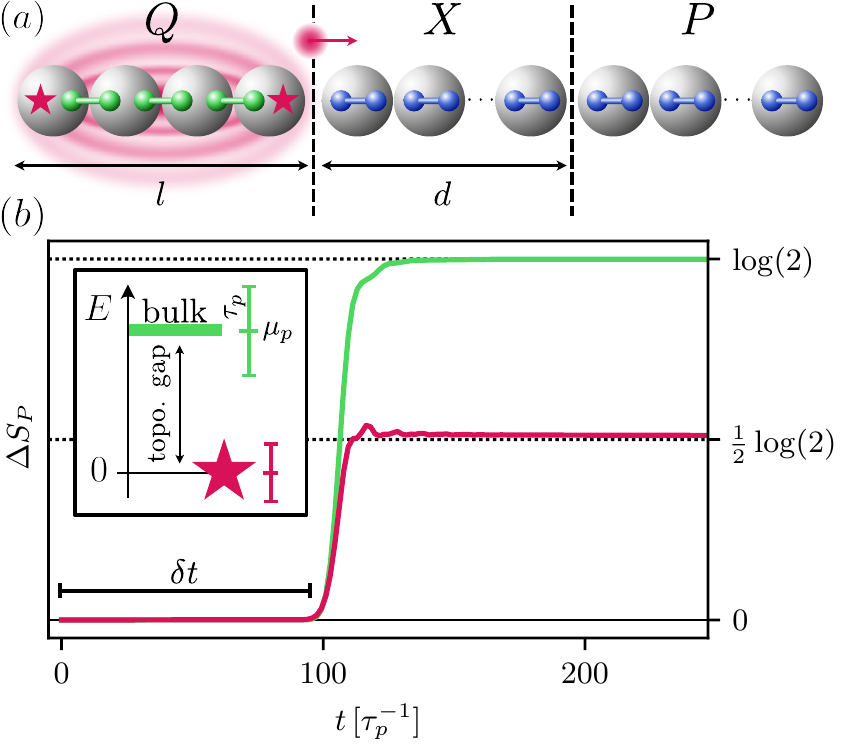}
	\caption{$(a)$ Hybrid quench-probe setup.
		A 1D Kitaev chain $Q$, which undergoes a quantum quench across its topological phase transition, is tunnel-coupled at its right end to the trivial regions $X$ and $P$. Quench-induced excitations are selectively transmitted through $X$ and eventually reach $P$, whose time-dependent entanglement properties are monitored. $(b)$ Quantized jumps in the entanglement entropy (EE) of the probe.
		When the latter is selectively coupled to the right Majorana zero mode (MZM) (red star), a robust fractional increase of the EE $\Delta S_P = \log(2)/2$ is observed (red line), which is half of the increase observed when the probe is coupled to the fermionic bulk flatband of the Kitaev chain (green line).
		A schematic of the energy-selective coupling, allowed by the quench probe approach, is provided in the inset. For our choice of parameters see \bibnotemark{}.}
		\bibnotetext{We choose the following parameter $(\mu^i,\tau^i = \Delta^i=\tau_p)\rightarrow (\mu^f =0, \tau^f=\Delta^f=11.76 \tau_p); N=500,l=4,d=100,t_0=10,\tau_t=1 \tau_p$.}
	\label{fig:SysSetup}
\end{figure}%
As a powerful and genuinely quantum-mechanical diagnostic tool, the time-evolution of entanglement has been widely investigated \cite{Gogolin2016,Calabrese2016b,Essler2016,Calabrese2020notes}, including dynamical signatures of topology such as protected crossings in the entanglement spectrum \cite{Gong2018,Pastori2020,PhysRevResearch.3.033022, 10.21468/SciPostPhysCore.3.2.012,Mondal2022}.
In homogeneous integrable systems, the spreading of entanglement after a quench is closely related to the propagation of pairs of entangled quasiparticle excitations with opposite momenta \cite{Calabrese2005,Calabrese2016,Kim2013,Calabrese2020notes}.
For more complex scenarios, involving for example periodic spatial modulations \cite{Bastianello2018} or open systems \cite{PhysRevB.103.L020302,Maity2020},  richer entanglement structures related to quench-induced excitations represent a  frontier of ongoing research \cite{PhysRevB.90.075144,PhysRevB.97.245135,BertiniFagottiPiroliCalabreseEntanglementEvolutionAndGeneralisedHydrodynamicsNoninteractingSystems,10.21468/SciPostPhys.7.1.005,PhysRevB.99.045150,Bertini2017,Bertini2018,Bastianello2020}.

In this Letter, we propose a novel approach for the study of entanglement dynamics in complex systems to selectively analyze a subset of quench-induced excitations.
This enables us to identify unique features in the spreading of entanglement, such as fractional jumps of the entanglement entropy (EE).
We directly relate their presence to the existence of non-trivial eigenmodes in the post-quench Hamiltonian, e.g. topological localized modes.
This remarkable capability stems from the hybrid nature of our proposed setup, sketched in Fig.~\ref{fig:SysSetup}(a), where only a part of the system ($Q$) is quenched while the entanglement is measured in a different (static) region ($P$), the two being coupled via a (static) separation layer ($X$).
This quench-probe approach provides a new perspective for the analysis of entanglement spreading in highly inhomogeneous systems, paving the way for novel observations that complement the study of (almost) homogeneous setups \cite{Calabrese2005,Calabrese2016,Eckardt2017,Cooper2019,Alba2017,Alba2018,DiGiulio2019,Bastianello2018,Bastianello2020}.
Its energy selectivity -- due to resonant coupling between regions $Q$ and $P$ -- is reminiscent of scanning tunneling spectroscopy.

As a specific case study, we use our approach to analyze the dynamics of the entanglement generated by a localized Majorana zero mode (MZM), hosted by a Kitaev chain (KC) \cite{Majorana1937,Kitaev2001,Alicea2012}.
This leads us to the discovery of quantized jumps in the EE of the probe with fractional amplitude
\begin{equation}
\Delta S_P = \frac{\log(2)}{2}.
\end{equation}
The corresponding trace is provided by the red line in Fig.~\ref{fig:SysSetup}(b).
Such a fractional increase, associated with the fractional entropy of a single MZM \cite{Sela2019,Smirnov2015,Silva2020}, clearly differs from the conventional EE increase $\Delta S_P = \log(2)$ that originates from an ordinary fermionic mode [see the green line in Fig.~\ref{fig:SysSetup}(b)].
The quantization is robust with respect to parameter variations but highly sensitive to the hybridization of two MZMs. 
These findings, representing a novel dynamical signature associated with a truly isolated MZM, are corroborated by the additional analysis of the mutual information (MI) shared between $Q$ and $P$ \cite{Alba2018,Maity2020,Mondal2022}, which allows us to identify spurious contributions to the EE and highlight the fractional entanglement jumps.
Importantly, the observation of this toplogical signature only requires the post-quench Hamiltonian to be topological, while the system can be prepared in a trivial thermal state.
The topological nature (and robustness) of an isolated MZM is the origin of the fractional value of $\Delta S_P$. Our setup is applicable to a variety of systems with particular entanglement spreading of either bulk or edge modes. Due to energy selective coupling, we are able to single out the contributions from a subset of modes, if they are separated in energy.\\
\indent\textit{Hybrid quench-probe setup.\textemdash} 
We consider the system depicted in Fig.~\ref{fig:SysSetup}(a), consisting of the three parts labeled $Q,X,$ and $P$. 
The first one, $Q$, is the one eventually undergoing a quantum quench. It is an $l$-site KC described by the Hamiltonian
	\begin{align}
		H^{Q}= \mu \sum_{i=1}^{l} c_i^\dagger c_i + \sum_{i=1}^{l-1} \left( \frac{\tau}{2} c_i^\dagger c_{i+1} +  \frac{\Delta}{2} c_i c_{i+1}+\text{h.c.}\right).
		\label{eq:HamKitaevChain}
	\end{align}
The operators $c_i^\dagger$ $(c_i)$ create (annihilate) a spinless fermion at site $i$, $\mu$ is the chemical potential, $\tau$ the nearest-neighbor hopping amplitude and $\Delta$ the superconducting pairing amplitude. For simplicity, we consider those parameters to be non-negative real numbers.
The KC features two different gapped phases, a trivial one for $|\mu|>\tau$ and a topological one for finite $\Delta$ and $|\mu|<\tau$.
At the topological sweet spot (TSS), i.e. $\tau=\Delta$ and $\mu=0$, the analysis of $H^Q$ in terms of Majorana operators $c_j = \frac{1}{2} (i \gamma_{2i-1}+\gamma_{2i})$ reveals the presence of two completely isolated MZMs at the two open ends of the chain $[\gamma_1,H^{Q}]=[\gamma_{2l},H^{Q}]=0$, depicted by red stars in Fig.~\ref{fig:SysSetup}(a).
The bulk of the KC at the TSS is described by a flat band at finite energy $E^Q = \tau$.
Deviations from the TSS (within the topological phase) imply an exponential leakage of the MZMs into the bulk, whose spectrum acquires then a finite bandwidth $E^Q(k) = \sqrt{(\tau \cos(k) + \mu)^2+(\Delta \sin(k))^2}$ \cite{Alicea2012}.
For simplicity, we illustrate the main features of our setup at the TSS. However, the observation of fractional EE is not limited to the TSS as we show below.  

The remaining $N-l$ sites of the system are described by a tight-binding Hamiltonian 
\begin{align}
H^{XP} = \sum_{i=l+1}^N \mu_p c^\dagger_i c_i + \frac{1}{2} \sum_{j=l+1}^{N-1} \tau_p (c^\dagger_i c_{i+1} + h.c.),
\end{align} 
with chemical potential $\mu_p$ and hopping amplitude $\tau_p$.
The corresponding spectrum reads
 \begin{align}
	\label{eq:E^XP}
	E^{XP}(k) = \mu_p + \tau_p \cos(k).  
\end{align} 
The first $d$ sites, i.e. the ones between $l<j\leq l+d$, form the separation layer $X$, while the probe region $P$ consists of the remaining sites with $l+d< j\leq N$.
The presence of a finite $X$ allows us to consider regimes in which the probe region $P$ is exclusively affected by quench-induced excitations that propagate ballistically in the chain, filtering out possible contributions to the entanglement associated with the $Q$-$X$ interface.
Regions $Q$ and $X$ are connected via a standard tunneling Hamiltonian
 \begin{align}
 \label{eq:htunn}
 	H^{T} = \frac{\tau_t}{2} (c_l^\dagger c_{l+1} + h.c.) = \frac{\tau_t}{4} [(i\gamma_{2l-1}+ \gamma_{2l}) c_{l+1} + h.c.],
 \end{align}
with coupling strength $\tau_t$.

It is particularly instructive to express fermions in terms of the corresponding Majorana operators.
At the TSS, $\gamma_{2l}$ is an isolated MZM while $\gamma_{2l-1}$, together with $\gamma_{2l-2}$, belongs to an ordinary fermionic mode of the flat bulk band of the KC. Coupled Majorana operators belonging to the bulk of the KC are depicted by green circles in Fig.~\ref{fig:SysSetup}(a).
By properly tuning the parameters of the system, it is thus possible to define two separated regimes.
For $|\mu_p|<\tau_p\ll\tau$ the probe is exclusively coupled to the MZM at the right end of the topological KC. By contrast, for $|\mu_p|\sim\tau\gg\tau_p$, the probe is coupled to the bulk band \cite{[{See the supplemental material, which includes Refs.\cite{Levy2019,Vidal2003,Calabrese2004,Holzhey1994,Islam2015,Kaufman2016,Lukin2019,Brydges2019}. }] supp}.
A sketch of this energy-selective coupling is provided in the inset of Fig.~\ref{fig:SysSetup}(b).
The exploitation of energy and spatial sensitivity, together with the presence of a separation layer $X$, differentiates our proposal from other quench-probe scenarios, such as the ones discussed in \cite{Calzona2017,Calzona2018,Ruggiero2021}. 

\textit{Quench procedure.\textemdash} The quench of region $Q$ consists in the abrupt change, at $t=0$, of the parameters $(\mu^i,\tau^i,\Delta^i)\to(\mu^f,\tau^f,\Delta^f)$.
We assume the system to be initially prepared in the ground state $|\psi_0\rangle$ of the initial Hamiltonian $H^i =H^{Q}(\mu^i,\tau^i,\Delta^i) + H^{XP} + H^{T}$. For $t\geq 0$, the time evolution of the system is instead controlled by the final Hamiltonian $H^f =H^{Q}(\mu^f,\tau^f,\Delta^f) + H^{XP} + H^{T}$. With respect to $H^f$, the state $|\psi_0\rangle$ consists of several quasiparticle excitations, that are emitted in both directions from every site in the quenched region $Q$.
Those counter-propagating quasiparticles are entangled between each other. Their motion is responsible for spreading of correlations and entanglement within the system, bounded by the Lieb-Robinson limit \cite{Lieb1972}.
For a wide range of homogeneous systems, these quasiparticles are produced in uncorrelated pairs, each one consisting of two entangled quasiparticles with opposite momenta \cite{Cazalilla2012,Schuricht2012,Cazalilla2006,DeNardis2014,Brockmann2014}.
The physics is richer in presence of interactions and/or inhomogeneities, which can lead to the presence of quasiparticle multiplets and non-trivial correlations \cite{Bertini2017,Bertini2018,Bastianello2018,Bastianello2020}.
When $H_f$ is chosen in the topological regime, our system is spatially inhomogeneous due to the presence of a pair of isolated MZMs.
This observation naturally raises the question whether the quasiparticles originating from the MZMs differ from the ones associated with the fermionic bulk of the KC.
Our proposed quench-probe setup proves to be particularly effective in providing an affirmative answer to this question.
\begin{figure}
	\centering
	\includegraphics[width=\columnwidth]{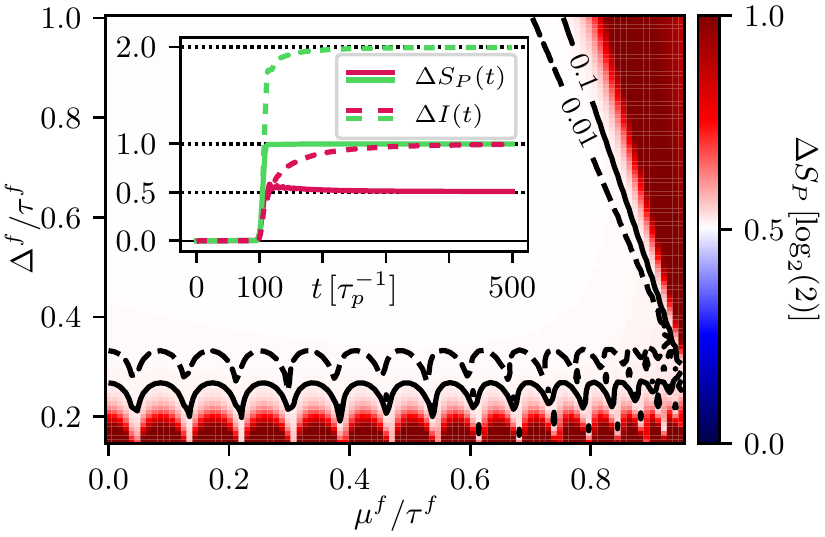}
	\caption{Fractional quantization of the EE increase as a function of $\mu^f/\tau^f,\Delta^f/\tau^f$ ratios. The solid and dotted black lines show the energy splitting of the MZMs in units of $10^{-3} \tau_p$. Inset: Quantized jumps in the EE $\Delta S_p(t)$ (solid lines) and MI $\Delta I(t)$ (dashed lines), in units of $\log_2(2)$ for a probe coupled either  to the MZM (red lines) or the fermionic bulk modes (green lines). To selectively couple to the MZM (fermionic bulk modes) we choose $\mu_p=0\, (\mu_p=\tau_f)$. For our choice of parameters see \bibnotemark{}.}
	\bibnotetext{The parameters used in the main figure are $(\mu^i=20\tau_p,\tau^i=\Delta^i=\tau_p,\tau^f=20 \tau_p); N=700,l=35,d=100,t_{sf}=1100 \tau_p^{-1},\tau_t=1 \tau_p$. In the inset our parameter choice is $(\mu^i,\tau^i = \Delta^i=\tau_p)\rightarrow (\mu^f =0, \tau^f=\Delta^f=20 \tau_p); N=500,l=4,d=100,t_0=10\tau_p^{-1},\tau_t=1 \tau_p$. Note, for ease of computing we use $l=4$ if we quench to the TSS where the MZMs are perfectly localized at the edges. For quenches away from the TSS we need to increase $l$ to avoid hybridization.}
	\label{fig:MIBulkEdge}
\end{figure}

\textit{Entanglement dynamics.\textemdash} The simplest way to analyze the entanglement properties of $P$ is to compute its EE, defined as 
\begin{align}
		S_P(t) = -\text{Tr} [\rho_P(t) \log(\rho_P(t))].
	\label{eq:EntEntropy}
\end{align}
Here, $\rho_P(t)$ is the reduced density matrix of the probe $\rho_P(t) = \text{Tr}_{QX} [\rho(t)]$, whose spectrum can be calculated from the single-particle correlation matrix  \cite{Peschel2003,Peschel2009,Vidal2003}.
The time-dependent variation of the EE after a quench to the TSS is shown in Fig.~\ref{fig:SysSetup}(b), where we plot $\Delta S_P(t) = S_P(t)-S_P(0)$ considering a selective coupling either to the isolated MZM (red line) or to the flat fermionic bulk band (green line).
After a finite time delay $\delta t$, we observe jumps in the EE that eventually reach either the trivial quantized value $\log(2)$ (for the coupling to the bulk) or an anomalous fractional value $\log(2)/2$ (for the coupling to the MZM).
Consistently with the quasiparticle picture, the time delay satisfies $\delta t \sim d \tau_p^{-1}$.
It can be interpreted as the time-of-flight associated with the excitations, emitted from the last site of $Q$, that propagates through the $d$ sites of $X$ at the maximum group velocity $\tau_p$ [see Eq.~\eqref{eq:E^XP}].
The lack of a steady linear increase of $\Delta S_P(t)$, typically observed in homogeneous systems \cite{Calabrese2005,Calabrese2016,Kim2013}, can be understood in terms of the vanishing group velocity in the bulk of the KC at the TSS.
This effectively freezes all the quasiparticles emitted in $Q$ with the only exception of the ones related to $\gamma_{2l-1}$ and $\gamma_{2l}$, which are directly connected to $X$ via $H^T$.
Those quasiparticles are ultimately responsible for the quantized jumps discussed before.

\begin{figure}
	\centering
	\includegraphics[width=\columnwidth]{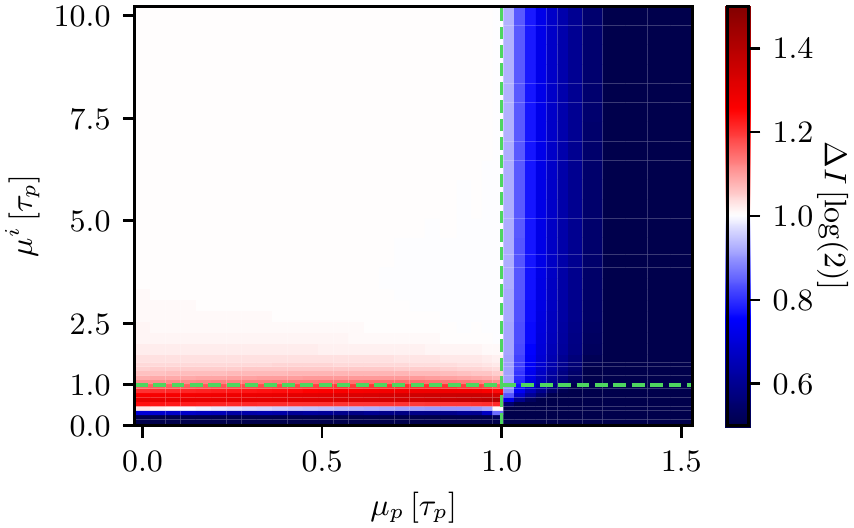}
	\caption{Quantization of the MI increase. $\Delta I$ is plotted as a function of $\mu^i$ and $\mu_p$ (units $\tau_p$). The dashed horizontal (vertical) line indicates the topological phase transition of the initial Hamiltonian (the transition between a gapless and gapped probe). For our choice of parameters see \bibnotemark{}.}
	\bibnotetext{The parameters used in the figure are $(\mu^i,\tau^i = \Delta^i=\tau_p)\rightarrow (\mu^f =0, \tau^f=\Delta^f=20 \tau_p); N=2000,l=4,d=100,t_0=80\tau_p^{-1},t_{sf}=3000\tau_p^{-1},\tau_t=1 \tau_p$.}
	\label{fig:MIStability}
\end{figure}

To strengthen the connection between the anomalous fractional jump of the EE and the presence of an isolated MZM, we additionally compute the time-dependent mutual information (MI) shared between the probe $P$ and the quenched region $Q$.
It is defined as
\begin{equation}
	I(t)= S_Q(t) + S_P(t) - S_{Q\cup P}(t)
\end{equation}
and quantifies the total amount of correlations between the two disjoint regions \cite{Alba2018,Maity2020,Mondal2022}, eliminating spurious contributions to $S_P$ coming from the separation layer $X$ and not from the quenched region $Q$. The increase of MI $\Delta I(t) = I(t)-I(t_0)$, where $t_0\lesssim \delta t$ and $I(t_0)\to 0$ for large $d$ \cite{supp}, is plotted in the inset of Fig.~\ref{fig:MIBulkEdge} (dashed lines).
It shares its main features with $\Delta S_P$.
In particular, when the probe is effectively coupled to the fermionic bulk of the KC (dashed lines), $\Delta I$ saturates at $2\log(2)$, indicating that $P$ and $Q$ share a conventional fermionic mode \cite{supp}. In contrast, when the probe is coupled to the isolated MZM (solid lines), the height of the increase is halved and $\Delta I$ saturates at $\log(2)$.
In the following, we carefully analyze the MZM case.

\begin{figure}
	\centering
	\includegraphics[width=\columnwidth]{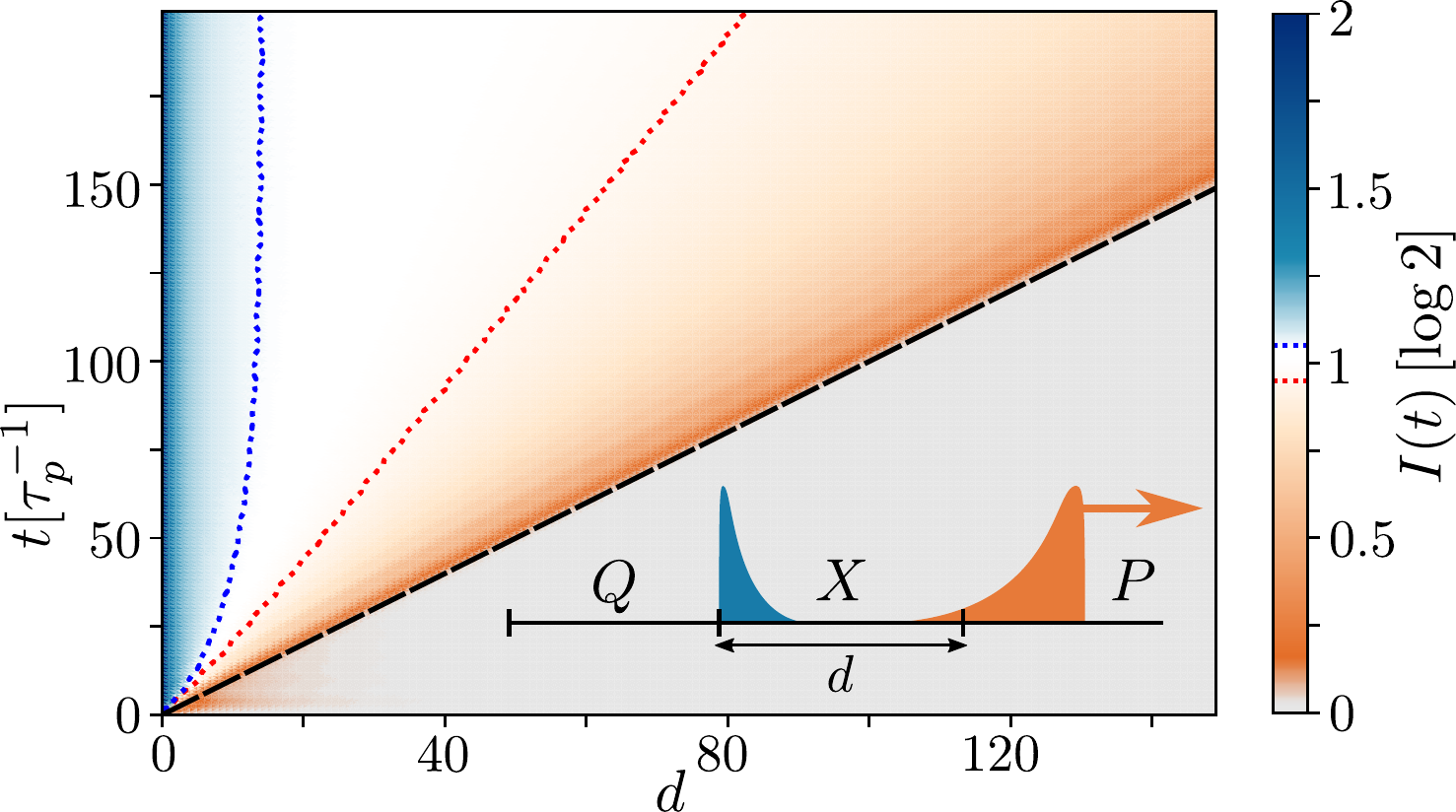}
	\caption{MI as a function of $d$ and time $t$. The red(blue)-dotted line corresponds to $I/\log2=0.95$ ($1.05$). The black-dashed line shows the Lieb-Robinson limit $d=t\tau_p$. The inset shows a sketch of the fixed (blue) and propagating (orange) contributions to the MI. For our choice of parameters see \bibnotemark{}.}
		\bibnotetext{We choose the following parameter $(\mu^i,\tau^i = \Delta^i=\tau_p)\rightarrow (\mu^f =0, \tau^f=\Delta^f=11.76 \tau_p); N=350,l=100,d=100,t_0=10,\tau_t=1 \tau_p$.}
	\label{fig:MITransient}
\end{figure}

\textit{Anomalous quantization.\textemdash} After a sufficiently long time $t_{sf}$ and in the large $d$ limit \footnote{Note that the maximum time that can be actually studied is limited by the finite number of sites $N$ because we want to avoid a signal that stems from the reflections of quasiparticles at the right end of the region $P$.
The dependence of the signal on $d$, the length of the region $X$, is further discussed in the description of Fig.~\ref{fig:MITransient}.}, $\Delta I(t_{sf})$ shows a high degree of quantization and robustness.
Indeed, as long as the probe is gapless and the initial Hamiltonian features a large trivial gap (such that regions $Q$ and $X$ are initially decoupled), the MI saturates at $\Delta I(t_{sf})=\log(2)$ without the need of fine-tuning, as shown by the extended white area in Fig.~\ref{fig:MIStability}.
Likewise, no fine-tuning of the tunnel coupling between $Q$ and $X$ is necessary to produce the anomalous quantization signature, as long as it is comparable to $\tau_p$  \cite{supp}.
This anomalous quantization is robust against finite temperature effects and deviations of $H^f$ from the TSS, as can be seen from the large white area in Fig.~\ref{fig:MIBulkEdge}.
Away from the TSS, two main effects matter:
(i) hybridization of MZMs and (ii) finite band-width of the fermionic bulk band.
Related to point (i), the hybridization of the MZMs disturbs the saturation of the EE at the fractional value of log(2)/2.
This makes sense because hybridized MZMs become regular fermions.
If the region Q is, however, chosen long enough such that the hybridization between the MZMs is weak, then the fractional EE can be observed, see Fig.2.
Related to point (ii), as long as the MZMs are energetically decoupled from the bulk, our quench-probe setup allows to isolate their contribution to the EE by energy selectivity.

As for the robustness at finite temperature, we show that the quantization of the MI is retained even when the system is initialized in a thermal trivial state of $H^i$ at finite temperature $T$, as long as the latter remains smaller than the topological gap $T\ll \tau^f=\Delta^f$ \cite{supp}. 


\textit{Non-equilibrium dynamics.\textemdash} Our quench-probe setup features another useful knob, i.e. the size $d$ of region $X$, which can significantly enrich the analysis of the post-quench entanglement dynamics.
Indeed, a careful study of $I(t)$ as a function of $d$ (plotted in Fig.~\ref{fig:MITransient}) reveals the coexistence of a fixed and a propagating component of the correlations contributing to the MI.
Let us explain this point by carefully inspecting Fig.~\ref{fig:MITransient}, from right to left.
If the probe region is too far away from $Q$, the quench-induced excitations have not yet reached $P$ and the MI is basically zero.
This explains the large triangular gray area in Fig.~\ref{fig:MITransient}, which is bounded by the Lieb-Robinson limit $d=t\tau_p$ (black dashed line) \footnote{The small deviations from $I=0$, which can be observed below the dashed line for small $d$, are a consequence of the non-local nature of the MI.}.
To the left of the Lieb-Robinson line, the MI increases as the probe includes an increasing number of sites entangled with $Q$ (orange region).
The MI reaches $I\simeq0.95\, \log(2)$ close to the red-dotted line, which we attribute to the propagation of excitations of finite but smaller group velocity than $\tau_p$.
To the left of the red-dotted line, the MI features a plateau around the anomalous quantized value of $\log(2)$ (white region), the regime described in the previous paragraphs.
For small $d$, i.e. when the probe region starts to include sites close to the $Q$-$X$ interface, the MI increases again and displays values above $\log(2)$ (blue region).
Interestingly, the correlations responsible for this additional increase of MI do not propagate within the probe, as shown by the blue-dotted line, corresponding to $I\simeq 1.05 \log(2)$, which is asymptotically vertical.
Finally, for $d=0$, the MI reaches the conventional quantized value of $I=2\log(2)$.
At a given (large) time, we can thus identify two groups of sites that are entangled with $Q$, a propagating one and a fixed one (pinned at the $Q$-$X$ interface), as sketched in the inset of Fig.~\ref{fig:MITransient}.
The precise and robust quantization of $\Delta I$, shown in Fig.~\ref{fig:MIStability}, can therefore be understood as the result of a dynamical phenomenon, namely the separation of the correlations between $Q$ and $P$ into two different components. 

\textit{Conclusions.\textemdash}
Our quench-probe setup allows us to identify a robust dynamical effect associated with the presence of an isolated MZM, hosted by the post-quench topological Hamiltonian.
The observation of this effect, consisting of particular fractional quantized jumps in the entanglement properties of the probe, only requires the preparation of the system in a trivial state.
Recent experimental progress shows that it is feasible to measure the second order Renyi entropy by quantum interference \cite{Islam2015,Kaufman2016,Lukin2019} or randomized measurements \cite{Brydges2019}. 
Even though the main focus of our work is put on the von Neumann entropy, we show in the SM \cite{[{See the supplemental material, which includes Refs. \cite{Levy2019,Vidal2003,Calabrese2004,Holzhey1994,Islam2015,Kaufman2016,Lukin2019,Brydges2019} }] supp} that all fractional features of the entanglement dynamics of MZMs can also be identified in the second order Renyi entropy.
Hence, we are confident that our quench-probe setup can be realized in engineered quantum systems similar to the ones discussed in Refs. \cite{Islam2015,Kaufman2016,Lukin2019,Brydges2019}.\\

\begin{acknowledgments}
	This work was supported by the W\"urzburg-Dresden Cluster of Excellence ct.qmat, EXC2147, project-id 390858490, and the DFG (SPP 1666 and SFB 1170). We thank the Bavarian Ministry of Economic Affairs, Regional Development and Energy for financial support within the High-Tech Agenda Project “Bausteine für das Quanten Computing auf Basis topologischer Materialen”.
\end{acknowledgments}

\bibliography{Arxiv-Resubmisson.bbl}

\begin{thebibliography}{79}
\bibitem [{2()}]{2}%
  \BibitemOpen
  \href@noop {} {}\bibinfo {note} {We choose the following parameter $(\mu
  ^i,\tau ^i = \Delta ^i=\tau _p)\rightarrow (\mu ^f =0, \tau ^f=\Delta
  ^f=11.76 \tau _p); N=500,l=4,d=100,t_0=10,\tau _t=1 \tau _p$.}\BibitemShut
  {Stop}%
\bibitem [{4()}]{4}%
  \BibitemOpen
  \href@noop {} {}\bibinfo {note} {The parameters used in the main figure are
  $(\mu ^i=20\tau _p,\tau ^i=\Delta ^i=\tau _p,\tau ^f=20 \tau _p);
  N=700,l=35,d=100,t_{sf}=1100 \tau _p^{-1},\tau _t=1 \tau _p$. In the inset
  our parameter choice is $(\mu ^i,\tau ^i = \Delta ^i=\tau _p)\rightarrow (\mu
  ^f =0, \tau ^f=\Delta ^f=20 \tau _p); N=500,l=4,d=100,t_0=10\tau _p^{-1},\tau
  _t=1 \tau _p$. Note, for ease of computing we use $l=4$ if we quench to the
  TSS where the MZMs are perfectly localized at the edges. For quenches away
  from the TSS we need to increase $l$ to avoid hybridization.}\BibitemShut
  {Stop}%
\bibitem [{8()}]{8}%
  \BibitemOpen
  \href@noop {} {}\bibinfo {note} {We choose the following parameter $(\mu
  ^i,\tau ^i = \Delta ^i=\tau _p)\rightarrow (\mu ^f =0, \tau ^f=\Delta
  ^f=11.76 \tau _p); N=350,l=100,d=100,t_0=10,\tau _t=1 \tau _p$.}\BibitemShut
  {Stop}%
\end{thebibliography}
\onecolumngrid
\newpage
\setcounter{secnumdepth}{2}
\appendix
\renewcommand{\thesection}{\Alph{section}}
\renewcommand{\thesubsection}{\Alph{section}\arabic{subsection}}
\renewcommand{\thefigure}{\Alph{section}\arabic{figure}}
\renewcommand{\theequation}{\Alph{section}\arabic{subsection}.\arabic{equation}}
\numberwithin{equation}{section}
\renewcommand{\thetable}{\Alph{section}\arabic{subsection}.\arabic{table}}
\numberwithin{table}{subsection}
\makeatletter
\renewcommand{\p@subsection}{}
\makeatother
\section*{Appendix}
\noindent In the main text, we propose a novel quench-probe setup for the study of entanglement dynamics and demonstrate its capabilities by showing a fractional entanglement signature induced by a single Majorana zero mode.
The main goal of this supplemental material is to provide additional technical details about origin and robustness of this signature.
This includes a discussion about the origin of the fractional entanglement signature in the mutual information (Section~\ref{sec:FractionalMajo}), an analysis of the initial entanglement (Section~\ref{sec:IniValueEEMI}), a study of the dispersion of the entanglement quasiparticles in the $X$ region (Section~\ref{sec:DispersionX}), an extended analysis of the robustness of the fractional entanglement signature (Section~\ref{sec:RobustnessAnalysis}), and an analysis of the fractional entanglement signature using the experimental relevant second order R\'{e}nyi entropy (Section~\ref{sec:RenyiEntropy}).
\section{Majornana-related MI}
\label{sec:FractionalMajo}
\setcounter{figure}{0} 
The goal of this section is to compute the entanglement entropy (EE) and the mutual information (MI) for subsystems that share either an ordinary fermion or a single Majorana.
To this end, we consider a toy model consisting of three physical fermionic sites whose creation (annihilation) operators read $c_j$ ($c_j^\dagger$) with $j=1,2,3$.
In the following, we use the eigenvalues of the number operators $n_j=c^\dagger_jc_j$ to label the states as $|n_1n_2n_3\rangle$.
As discussed in the main text, it is convenient to describe the system in terms of Majorana operators $\gamma_j$ that satisfy
\begin{equation}
\label{eq:app:c-gamma}
	\begin{cases}
		c_j = \tfrac{1}{2} (i\gamma_{2j-1}+\gamma_{2j}) \\
		c_j^\dagger = \tfrac{1}{2} (-i\gamma_{2j-1}+\gamma_{2j}) \\
	\end{cases}
\Leftrightarrow \;
	\begin{cases}
		\gamma_{2j} = c_j+c_j^\dagger\\
			\gamma_{2j-1} = i(c_j^\dagger-c_j)\\
	\end{cases}.
\end{equation} 

We start our analysis by considering a pure state of the system,
\begin{equation}
|\psi_F\rangle = \frac{|100\rangle+|010\rangle}{\sqrt{2}}, 
\end{equation}
which is an eigenstate $F|\psi_F\rangle = |\psi_F\rangle$ of the Hermitian operator $F = c_1^\dagger c_2 +c_2^\dagger c_1$.
As the latter describes the tunneling of one fermion between the first two sites, $|\psi_F\rangle$ features a delocalized fermion distributed over the first two sites. 
A straightforward analysis of the entanglement entropies associated to every single site leads to $S_1^F=S_2^F=\log(2)$ and $S_3^F=S_{1\cup 2}^F=0$.
The MI between the first and the second site is thus given by 
\begin{equation}
I^F = S_1^F+S_2^F-	S_{1\cup 2}^F = 2\log(2).
\end{equation}

The expression of the operator $F = (i\gamma_2\gamma_3 - i\gamma_1\gamma_4)/2$ in terms of Majorana operators shows that a fermionic tunneling actually corresponds to two Majorana tunneling terms, expressed via the parity operators $P_{ij}=i\gamma_i\gamma_j$ (remember that $\gamma_i^\dagger =\gamma_i$).
This raises the question about the entanglement properties of the system when it is in an eigenstate of only a \textit{single} Majorana tunneling term, say $P_{23}$. To this end, we consider the state
\begin{equation}
|\psi_M\rangle = \frac{|100\rangle+|010\rangle+|001\rangle+|111\rangle}{2}
\end{equation}
that indeed satisfies 
\begin{equation}
	\begin{split}
	P_{23}|\psi_M\rangle &= (F+ c_1 c_2 + c_2^\dagger c_1^\dagger) |\psi_M\rangle \\
	&= \frac{|010\rangle+|100\rangle+|111\rangle+|001\rangle}{2} =  |\psi_M\rangle.
\end{split}
\end{equation}
However, $|\psi_M \rangle$ is not an eigenstate of the second Majorana tunneling term 
\begin{equation}
\langle \psi_M |P_{14}|\psi_M\rangle = 0.
\end{equation}
The single-site reduced density matrices obtained from $|\psi_M\rangle $ are all maximally mixed. This leads to $S^M_1 = S^M_2 = S^M_3 = S^M_{1\cup 2} = \log(2)$. The MI between the first and the second site is thus given by
\begin{equation}
I^M = S_1^M+S_2^M-	S_{1\cup 2}^M = \log(2).
\end{equation}

To summarize, when two sites equally share a fermion, their MI reads $I^F = 2\log(2)$. By contrast, when they only share one Majorana, the MI is halved, i.e. $I^M = \log(2)$. 
\section{Initial entanglement value of EE and MI}
\label{sec:IniValueEEMI}
\setcounter{figure}{0}
In the main text (see Figs.~1-3), we focus on the entanglement entropy of the probe $\Delta S_P(t)= S_P(t)-S_P(0)$ and the mutual information $\Delta I(t) = I(t) - I(t_0)$ between the $Q$ and $P$ region, as only the change of those measures is important to observe the fractional entanglement value of the MZM.
The aim of this section is to discuss more thoroughly the genuine measures $S_P(t)$ and $I(t)$ since both exhibit a finite value for times $0\leq t < \delta t$ before the delay time $\delta t$. 
Here, $\delta t$ is the time period between the quench and the first entanglement quasiparticles arriving in region $P$, which can be identified by a rapid increase of the entanglement measure.
The origin of this initial value is completely different between the EE and MI, so we discuss them separately in the following.\\
\subsection{Initial entanglement entropy}
\begin{figure}
\centering
\begin{minipage}[t]{.49\textwidth}
  \centering
	\includegraphics[width=\textwidth]{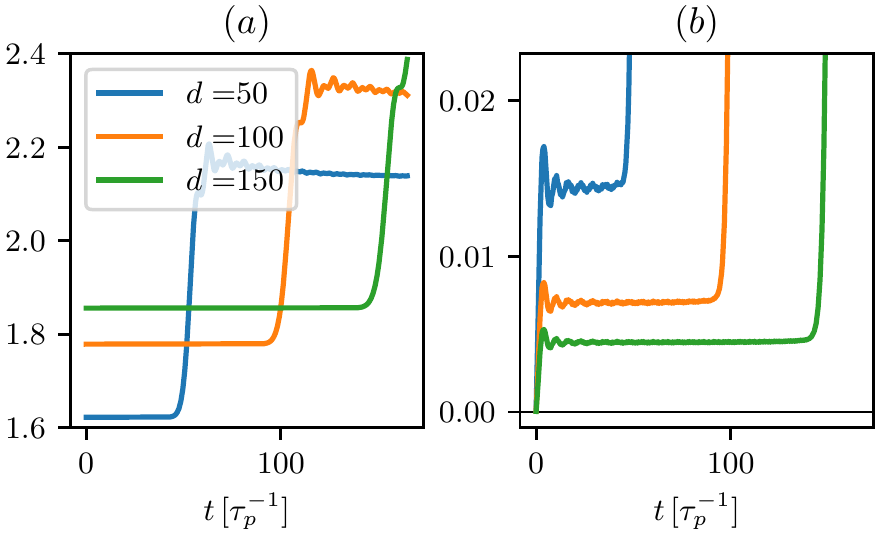}
	\caption{Illustration of the initial offset of $(a)$ the EE and $(b)$ the MI of the probe for different sizes $d$ of the spacing region $X$ when coupled to the MZM. We show the amount of entanglement in units of $\log(2)$ with $N=500,l=4,\tau_t=1\tau_p; (\mu^i=20,\tau^i=\Delta^i=1\tau_p)\rightarrow (\mu^f=0,\tau^f=\Delta^f=20\tau_p)$ }
	\label{fig:SM-IniOffset}
\end{minipage}\hfill
\begin{minipage}[t]{.49\textwidth}
  \centering
	\includegraphics[width=\textwidth]{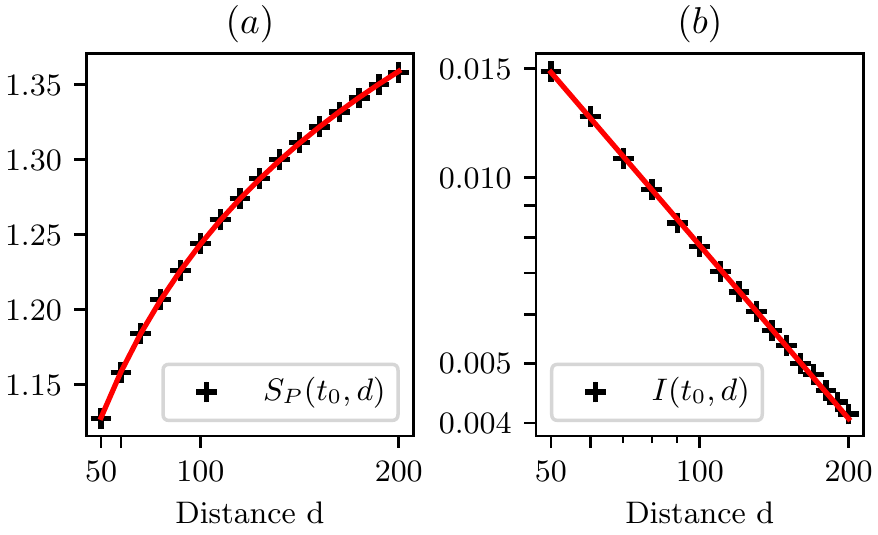}
	\caption{Analysis of initial offset of $(a)$ the EE $S_P(t_0,d)$ and $(b)$ the mutual information $I(t_0,d)$ with respect to $d$, when coupled to the MZM. The MI is given in units of $\log(2)$, the EE in natural units. For $(a)$ we choose $t_0=0$.  As a guide to the eye a power law (red line) is fitted to the data, for the MI $I(t_0,d)\propto d^{-0.93}$. For the EE we use a logarithm to fit the data $S_P(t_0,d)\propto \log(d)/6.0+ \text{const.}$, which is the standard result known from literature \cite{Levy2019}. In $(b)$ we numerically determine $I(t_0)$ as the value of $I(t)$ $(t\approx \delta t)$ at the time $t_0(d)$ when the slope of the mutual information is above the threshold of $\frac{d}{dt}I(t)>0.0005$. Due to oscillations, we smooth the data using a Savitzky-Golay filter before taking the numerical derivative.  Our parameter choice is $(a): N=1000; (b): N=2000,t_0=0; l=4,\tau_t=1\tau_p; (\mu^i=20,\tau^i=\Delta^i=1\tau_p)\rightarrow (\mu^f=0,\tau^f=\Delta^f=20\tau_p)$.}
	\label{fig:SM-IniOffset-dDependence}
\end{minipage}
\end{figure}

In Fig.~\ref{fig:SM-IniOffset}$(a)$, we study $S_P(t)$ of the probe, when coupled to the MZM, from $t=0$ to approximately $t\sim \delta t$ for several setups with varying distances $d$.
Here, the important observation is that $S_P(t)$ has a constant finite value from $t=0$ until $t\sim \delta t$ and which is similar in magnitude as the height of the (fractionalized) jump of $\log(2)/2$.
The finite constant EE is due to the fact that we use a gapless probe.
It is well known that a subset (here the $P$-region) of a gapless system (here the $XP$-region) has a non-zero EE \cite{Holzhey1994}.
The initial value vanishes as soon as the probe becomes gapped.\\
It is expected that, for a gapless probe, the initial EE grows logarithmically with the size $d$ of the $X$-region.
We verify this in Fig.~\ref{fig:SM-IniOffset-dDependence}$(a)$, where we plot $S(t_0,d)$ for several distances $d$ and fit a logarithmic function (red curve) to the data points.
It follows $S_p(t_0,d)\propto \log(d)/6 +\text{const.}$, in agreement with the expectation for a 1D fermionic system with a single entanglement cut and a central charge $c=1$ \cite{Levy2019,Vidal2003,Calabrese2004,Holzhey1994}.\\
\subsection{Initial mutual information}
The MI behaves completely different to the EE for times $0\leq t < \delta t$, as can be seen in Fig:~\ref{fig:SM-IniOffset}$(b)$, where we plot the MI when coupled to the MZM for several distances $d$.
Initially, at $t=0$, the MI is $I(t)\sim 0$ because the $Q$ and $P$ region are initially decoupled from each other.
Subsequently, we observe a time evolution consisting of damped oscillations, which eventually decay into a plateau at a small but finite value of $I$, see Fig.~\ref{fig:SM-IniOffset}$(b)$.
The height of the plateau $I(t_0)$ is typically much smaller (two orders of magnitude in Fig.~\ref{fig:SM-IniOffset}$(b)$) than the height of the jump occurring at $t \sim \delta t$.
In addition, the height of the plateau $I(t_0)$ goes to zero as $d$ increases (see Fig.~\ref{fig:SM-IniOffset}$(b)$ and Fig.~\ref{fig:SM-IniOffset-dDependence}$(b)$).
It follows a powerlaw decay (red line in Fig.~\ref{fig:SM-IniOffset-dDependence}$(b)$).
A non-vanishing (and time-dependent) MI before $\delta t$ might seem puzzling, given the fact that the information about the quench reaches region $P$ only after a time $\delta t$, as shown by the EE.
However, the reason is that the MI is not a local quantity but, by definition, it involves the EE of $Q$, $P$ and $X$.
Right after the quench, region $Q$ starts to be coupled only to the first sites of region $X$.
They, however, are already weakly correlated with region $P$, given the gapless nature of $H^{XP}$.
Those initial correlations are indeed expected to decay as a powerlaw with $d$ (i.e. the distance of the sites from the entanglement cut between region $X$ and $P$).
This explains the non zero value of $I(t)$ below the Lieb-Robinson line for small $d$, visible in Fig.~4 of the main text.
We emphasize that $I(t_0)$ is very small for most values of $d$, so that it is fair to consider $\Delta I(t) \sim I(t)$ in most scenarios. 
\section{Quasiparticle dispersion in $X$ region}
\label{sec:DispersionX}
\setcounter{figure}{0}
We now focus on the long-time behavior of the MI, detailing how the limit value $\log(2)$ is reached when we couple to the MZM.
For this reason, we define $f(t)=1-\Delta I(t) /\log(2)$, which measures the distance of the mutual information to $1$ in units of $\log(2)$.
In Fig.~\ref{fig:SM-DispersionX}, we plot $f(t)$ on a double logarithmic scale for several setups.
In Fig.~\ref{fig:SM-DispersionX}$(a)$, we vary the distance $d$, while in Fig.~\ref{fig:SM-DispersionX}$(b)$, we change the chemical potential $\mu_p$ in the $XP$ region.
Both have in common that during the delay time $t\leq \delta t$, when non of the entanglement particles emitted from $Q$ have reached the probe region $P$, $f(t)=1$ remains constant.
However, after some transient for $t>\delta t$ we observe a power law decay of $f(t)$ towards zero.\\
In the following, we study how $\alpha$ depends on the system parameters.
In particular, we show that it is independent of $d$ (Fig.~\ref{fig:SM-DispersionX}$(a)$) but, in general, it depends on the other parameters e.g. $\mu_p$ (see Fig.~\ref{fig:SM-DispersionX}$(b)$).\\
Neglecting some transient effects, we can give a rough description of the MI as:
\begin{align}
I(t,d)=
	\begin{cases} 
      0 & t\leq \delta t(d) \\
     \log(2) \left[ 1- \left(\frac{t}{\delta t(d)}\right)^{-\alpha}\right] & t > \delta t(d)
   \end{cases}
   \label{eq:SM-PiecewiseMI}
\end{align}
Here, $\delta t(d)$ and $\alpha$ depend on the system parameters we choose for a specific setup.
\begin{figure}
  \centering
	\includegraphics[width=0.5\textwidth]{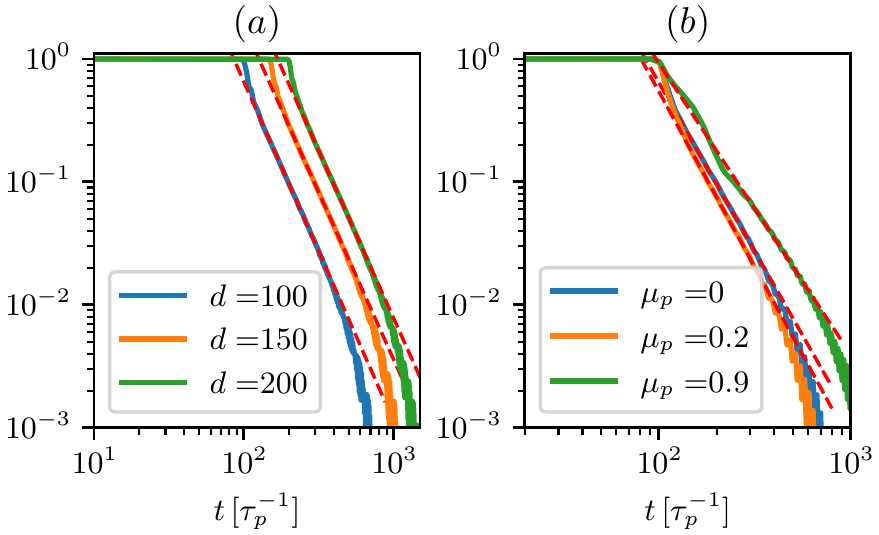}
	\caption{Illustrations of how varying $(a)$ the distance $d$ and $(b)$ the chemical potential $\mu_p$ affects the dispersion of the entanglement quasiparticles. For better visualization of the dispersion we plot the quantity $f(t)$ in units of $\log(2)$. There the red dashed lines show regions, where the decay follows a powerlaw. We choose $(a):N=1000;(b):N=800,d=100,l=4,\tau_t=1\tau_p,; (\mu^i=20,\tau^i=\Delta^i=1\tau_p)\rightarrow (\mu^f=0,\tau^f=\Delta^f=20\tau_p)$. For the plots $(a),(b)$ we adapt the distance $d$ respectively the potential $\mu_P$ (see legend entries).}
	\label{fig:SM-DispersionX}
\end{figure}%
The parameter $\alpha$ plays a particular role in the entanglement dynamics of our system as it determines the amount of correlation carried by modes of a specific velocity in the probe.
In the following, we want to define more precisely the relation between $\alpha$ and the aforementioned velocity.
Therefore, using Eq.~\eqref{eq:SM-PiecewiseMI}, let us determine the time $t_p$ needed to reach a certain level $p$ of MI, say $I(t_p)= p \log(2)$ (with $0<p<1$).
Then it follows that
\begin{align*}
	p&=1 -\left(\frac{t_p}{\delta t}\right)^{-\alpha}\\
	\Rightarrow 1-p &= \left(\frac{\delta t}{t_p}\right)^\alpha\\
	\Rightarrow t_p&=\delta t (1-p)^{-\frac{1}{\alpha}}
\end{align*}
with $\delta t = d/v_\text{max}$.
Hence, there is a linear relation between $t_p \propto d$, which we have already seen in Fig.~4 of the main text.
There, the red dotted line corresponds to the contour of $I(t_p)=p \log(2)$ with $p=0.95$.\\
From $t_p$, we can derive a velocity
\begin{align}
	v_p =\frac{d}{t_p}= v_\text{max} (1-p)^{\frac{1}{\alpha}}.
	\label{eq:SM-vP}
\end{align}
We can thus interpret Eq.~\eqref{eq:SM-vP} as the velocity of the modes whose arrival in the probe region is associated with $I$ reaching $I = p \log(2)$.
This picture is consistent with the idea that several quasiparticles are emitted and travel independently in the probe, with different velocities according to their momentum and the dispersion relation of the probe, each one carrying a contribution to the MI.
Large (small) values of $\alpha$ can be therefore associated with the fact that a large portion of the correlations is carried by fast (slow) modes.\\
That explains why varying $d$ in Fig.~\ref{fig:SM-DispersionX}$(a)$ has no effect on the decay of $f(t)$, since the distribution of entanglement carried by modes with different velocity is independent of the distance $d$ between the $Q$ and the $P$ region.
A change of $\mu_p$ (see Fig.~\ref{fig:SM-DispersionX}$(b)$) affects the aforementioned distribution of entanglement between the modes, leading to different decaying rates $\alpha$ of $f(t)$.
\section{Robustness and sensitivity of the fractional entanglement signature}
\label{sec:RobustnessAnalysis}
\setcounter{figure}{0}
In the main text, we explain that the fractional entanglement signature is robust against deviations from the TSS and finite temperatures $T$, while being very sensitive to hybridization of the MZMs. 
In the following we want to corroborate these statements.\\
\subsection{Sensitivity to hybridization} \label{sec:SensitivityHybridization}
\begin{figure}
\centering
\begin{minipage}[t]{.49\textwidth}
  \centering
	\includegraphics[width=\columnwidth]{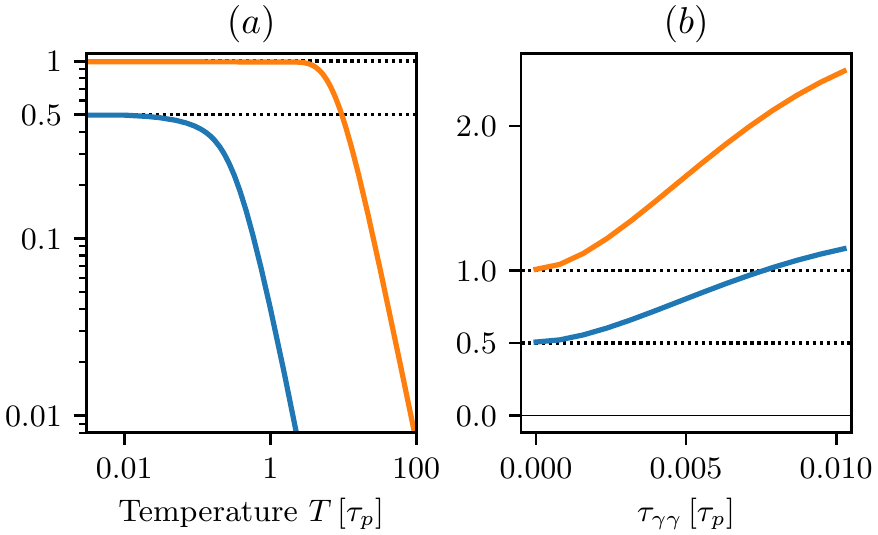}
	\caption{$\Delta I$ (orange lines) and $\Delta S_P$ (blue lines), both in units of $\log(2)$, as a function of temperature $T$ (panel $(a)$) and Majorana hybridization $\tau_{\gamma \gamma}$ (panel b) in the long time limit $t_{sf}$. We choose $l=4,d=100,\tau_t=1\tau_p,\tau _p=1;(\mu ^i=20,\tau^i=\Delta ^i=1\tau _p)\rightarrow (\mu ^f=0,\tau^f=\Delta ^f=20\tau _p)$ and $(a): N=500, t_{sf}=500; (b): N=2000,t_{sf}=3000$.}
	\label{fig:SM-MISensitivity}
\end{minipage}\hfill
\begin{minipage}[t]{.49\textwidth}
  \centering
	\includegraphics[width=\columnwidth]{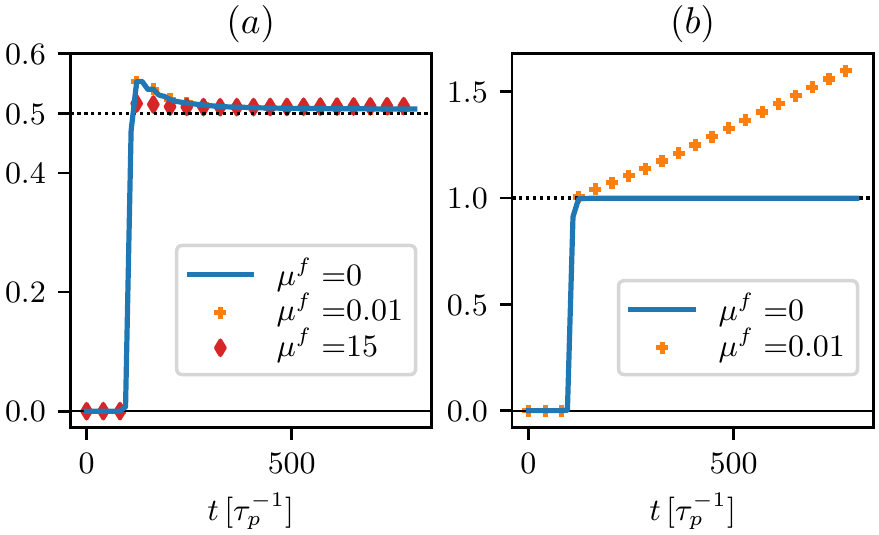}
	\caption{Effects on the time dependence of $\Delta S_P$ when the quench deviates from the TSS. The parameter $\mu^f$ (in units of $\tau_p$) is varied (see legend entries) and the time dependence of $\Delta S_P$ in units of $\log(2)$ is studied when coupled to $(a)$ the MZM respectively $(b)$ to the bulk states. Our choice of parameters is $N=1000,d=100,l=34,\tau_t=1\tau_p,; (\mu^i=20,\tau^i=\Delta^i=1\tau_p)\rightarrow (\mu^f,\tau^f=\Delta^f=20\tau_p)$.}
	\label{fig:SM-DeviationTSS}
\end{minipage}
\end{figure}

An important property of $\Delta I$ and $\Delta S_P$ is that they display a remarkable sensitivity to the presence of hybridization between the two MZMs hosted by the Kitaev chain (KC).\
To quantify this point, we consider an additional contribution to the Hamiltonian: 
\begin{align}
	H_{\gamma \gamma} = i \tau_{\gamma \gamma} \gamma_1 \gamma_{2l}+\text{h.c.}.
\end{align}
It directly couples the Majorana operators $\gamma_1$ and $\gamma_{2l}$ at the two ends of the KC  and a finite amplitude $\tau_{\gamma \gamma}>0$ leads to an hybridization of the two MZMs, which acquire a finite energy and cease to be isolated.
If we add $H_{\gamma \gamma}$ to our problem and stay otherwise at the topological sweet spot (TSS) then  $\tau_{\gamma \gamma}$ is the only parameter that controls the hybridization of the MZMs.
In this sense, it quantifies the hybridization.
At the same time, it mimics the situation away from the TSS in a finite length KC.
The sensitivity of the entanglement signature to hybridization can be seen in Fig.~\ref{fig:SM-MISensitivity}$(b)$, where we plot the values of $\Delta I(t)$ (orange line) and $\Delta S_P(t)$ (blue line) a long time $t=3\, 10^3\, \tau_p^{-1}$ after the quench as a function of $0\leq \tau_{\gamma \gamma} \leq \,10^{-2}\, \tau_p$.
Deviations from the quantized values can be observed already for $\tau_{\gamma \gamma} \sim 10^{-3} \tau_p$.
This remarkable sensitivity to Majorana hybridization strongly points at the isolated nature of the right MZM as the origin of the anomalous quantized jumps observed in both $\Delta S_P$ and $\Delta I$.
\\
\subsection{Robustness at finite temperature}
In sharp contrast to the high sensitivity with respect to Majorana hybridization, 
the anomalous quantization of $\Delta I=\log(2)$ is particularly robust with respect to other parameters.
In Fig.~\ref{fig:SM-MISensitivity}$(a)$, we study the effects of finite temperature $T$. 
This amounts to consider the system to be initialized not in the groundstate $|\psi_0\rangle$ of $H^i$, but rather in the thermal state $\rho_{th}(T)=\sum_m \exp(-\varepsilon_m/T) |\psi_m\rangle \langle \psi_m|$, where $|\psi_m\rangle, \varepsilon_m$ are eigenstates respectively eigenenergies of $H^i$ and $k_{\rm B}=1$.
After a quench to the TSS, the quantization of $\Delta I = \log(2)$ (orange line) is perfectly retained up to temperatures slightly higher than the probe bandwidth, $T\gtrsim \tau_p$, but smaller than the topological gap, $T\ll \tau^f=\Delta^f$.
This remarkable robustness at finite temperature strongly supports the topological origin of our phenomenon.
The EE is more sensitive to the presence of thermal correlations.
It displays a quantized jump in $\Delta S_P$ (blue line) only for $T\ll \tau_p$.\\
\subsection{Robustness against deviations from the TSS}
Importantly, the anomalous quantization of $\Delta I$ and $\Delta S_P$ is retained also when the final Hamiltonian $H^f$ is tuned away from the TSS (but still within the topological phase).
In Fig.~\ref{fig:SM-DeviationTSS}$(a)$, it is clearly visible, that for a sufficiently long $Q$ region $\Delta S_P$ converges to the fractional value of $\log(2)/2$ even for substantial deviations from the TSS (such as $\mu^f = 15 \tau_p$, which is of similar magnitude to $\tau^f=\Delta^f=20\tau_p$).
The same behavior can be observed in the MI.
\begin{figure}
  \centering
	\includegraphics[width=0.5\textwidth]{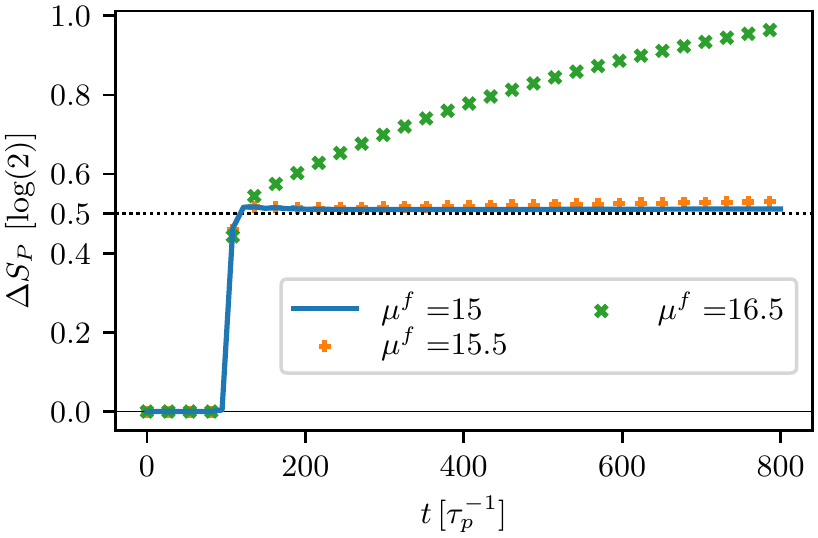}
	\caption{Effects of hybridization of the MZMs on the time dependence of $\Delta S_P$. The values of $\mu^f$ (in units of $\tau_p$) are chosen in such a way that at $\mu^f=15 \tau_p$ the hybridization of the Majoranas is $\Delta E \sim 0.001 \tau_p$, while for $\mu^f=16.5\tau_p$ $\Delta E \sim 0.018 \tau_p$. We choose $N=800,d=100,l=34,\tau_t=1\tau_p; (\mu^i=20,\tau^i=\Delta^i=1\tau_p)\rightarrow (\mu^f=0,\tau^f=\Delta^f=20\tau_p)$.}
	\label{fig:SM-DeviationTSS-Hybridization}
\end{figure}%
\\In Fig.~\ref{fig:SM-DeviationTSS-Hybridization}, we illustrate how the hybridization of the MZMs affects the time evolution of $\Delta S_P$ and the fractional entanglement signature.
For $\mu^f=15\tau_p$, the hybridization is very small, leading to an energy splitting between the MZM of $\Delta E\sim 10^{-3}\tau_p$.
As a result, according to Sec.~\ref{sec:SensitivityHybridization}, we expect the quantization to be retained until times of the order $t\sim 1000 \tau_p^{-1}$.
As $\mu^f$ increases, the hybridization becomes more sizable.
For $\mu^f=15.5 \tau_p$ (orange dots), the splitting is $\Delta E\sim 3 \cdot 10^{-3} \tau_p$ and, indeed, we start to see a deviation from the blue line around $t\sim 300\tau_p^{-1}$.
For $\mu^f=16.5 \tau_p$ (green), the splitting is even larger ($\Delta E \sim 0.018 \tau_p$) and $\Delta S_P(t)$ displays no fractional plateau anymore.\\
The situation is different when the probe is selectively coupled to the bulk of the KC.
Then, the finite group velocity of the bulk band allows for a large number of quench-generated quasiparticles above the topological gap to move from $Q$ to $P$ via $X$.
As a result, in the long time limit, $\Delta I$ and $\Delta S_P$ do not saturate anymore but rather increase linearly (before finite size effects associated with finite $l$ and $N$ matter), as illustrated in Fig.~\ref{fig:SM-DeviationTSS}$(b)$.
\subsection{Interplay of edge and bulk contributions}
\begin{figure}[H]
  \centering
  \begin{minipage}[t]{0.49\textwidth}
    \includegraphics[width=\textwidth]{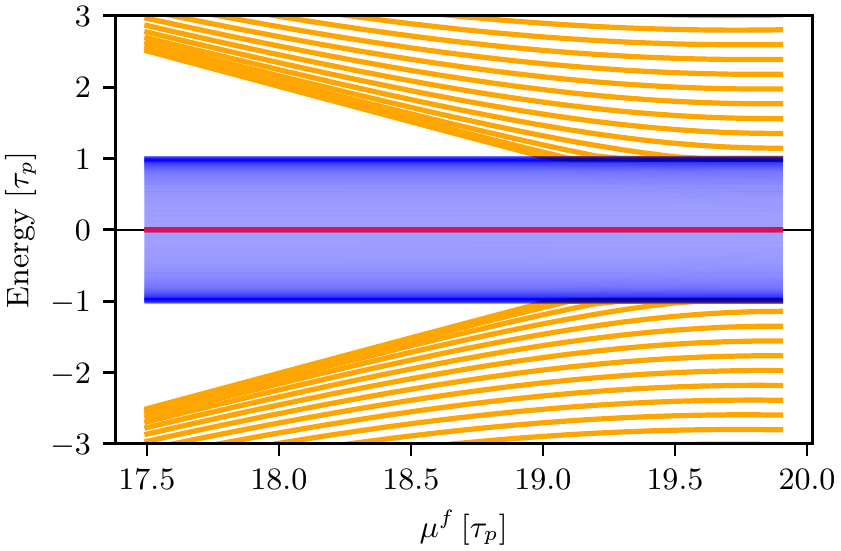}
    \caption{Spectrum of final Hamiltonian $H^f(\mu^f,\Delta^f=\tau^f=20 \tau_p); N=1000,l=301,\mu_p=0,\tau_t=1\tau_p,\tau_p=1$ for varying $\mu^f$. The MZMs are colored in red, the bulk modes of region $Q$ in orange and the dense modes in region $XP$ in blue.}
    \label{fig:SM-Spectrum-SmallBulkGap}
  \end{minipage}
  \hfill
  \begin{minipage}[t]{0.49\textwidth}
    \includegraphics[width=\textwidth]{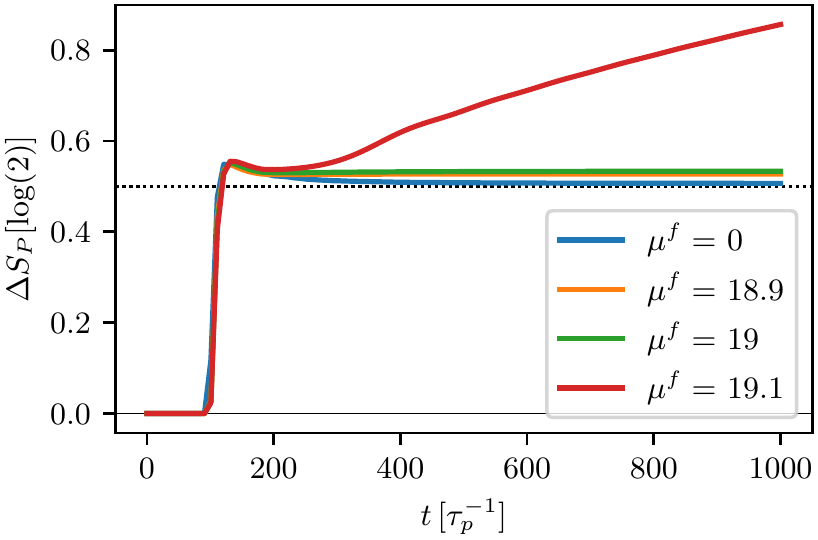}
    \caption{Entanglement entropy in the case of decreasing bulk gap $(\mu^i=20\tau_p,\Delta^i=0,\tau^i=1 \tau_p)\rightarrow (\mu^f=0,\Delta^f=\tau^f=20 \tau_p); N=1000,l=301,d=100,\mu_p=0,\tau_t=1\tau_p$}
    \label{fig:SM-EE-MZM-SmallBulkGap}
  \end{minipage}
\end{figure}

As long as the hybridization is small and the bulk gap sufficiently large the entanglement signature is perfectly fractional quantized. We now address the question whether we can still distinguish the edge contribution to the entanglement dynamics from the bulk contribution in case of a propagation of bulk modes into the probe.
This happens when the probe is coupled to the MZM and also partially to the bulk modes.
To clarify this point, we analyze the influence of bulk modes on the fractional signature of the edge modes.
Hence, we choose $\mu^f$ of the $Q$ region, such that the bulk gap is comparable to the bandwidth of the probe (see Fig.~\ref{fig:SM-Spectrum-SmallBulkGap}).
We clearly observe in Fig.~\ref{fig:SM-EE-MZM-SmallBulkGap} a sharp fractional jump due to the MZM on top of linear growth caused by the bulk modes.
This stresses again the significance of our result. The fractional entanglement signature of the MZM is recognizable in the presence of bulk modes.
\section{MZM entanglement signature in second order R\'{e}nyi entropy}
\label{sec:RenyiEntropy}
In our manuscript, we analyze the entanglement dynamics of a quenched Kitaev chain using a quench-probe setup.
There the MZM reveals a fractional signature in the time dependent von Neumann entropy.
In general, such entanglement signatures are challenging to detect experimentally.
However, impressive experimental progress has recently been made \cite{Islam2015,Kaufman2016,Lukin2019,Brydges2019}, where protocols aiming at observing the second order R\'{e}nyi entropy by using two "copies" of the system of interest respectively by means of randomized measurements have been employed.
In the following, we demonstrate that the main findings of our manuscript remain visible in the second order R\'{e}nyi entropy.
\subsection{R\'{e}nyi entropy signature of MZM}
\begin{figure}[H]
  \centering
    \includegraphics[width=0.5\textwidth]{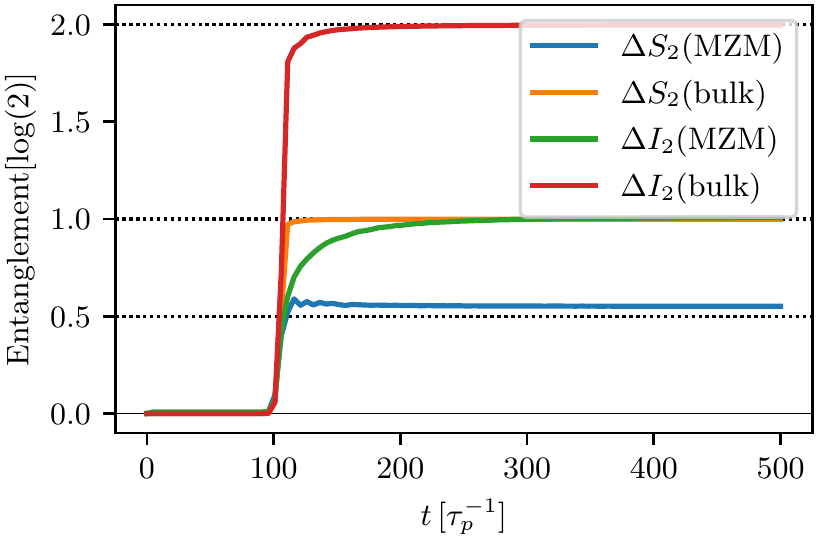}
    \caption{Quantum purity and R\'{e}nyi mutual information between the $Q$ and $P$ region for $q=2$ when coupled to the MZM respectively bulk modes. Quench parameters: $(\mu^i=20\tau_p,\Delta^i=\tau^i=1 \tau_p)\rightarrow (\mu^f=0,\Delta^f=\tau^f=20 \tau_p); N=500,l=4,d=100,\mu_p=0 \,(\text{MZM});\mu_p=20 \tau_p\,(\text{bulk})$}
    \label{fig:RFA-Renyi-Collection}
\end{figure}

The Renyi entropy of a subsystem $\rho_A$ is defined as
\begin{align}
	S_q = \frac{1}{1-q} \log(\text{Tr}[\rho_A^q]),
	\label{eq:RenyiEntropy}
\end{align}
hence the second order R\'{e}nyi entropy, also called quantum purity, corresponds to
\begin{align*}
	S_2 = - \log(\text{Tr}[\rho_A^2]).
\end{align*}
In the limit $q\rightarrow 1$, the Renyi entropy transforms to the standard von Neumann entropy.\\
Based on Eq.~\ref{eq:RenyiEntropy}, we can define the R\'{e}nyi mutual information (RMI) as
\begin{align}
	I_q(A,B) = S_q(\rho_A)+S_q(\rho_B)-S_q(\rho_{A\cup B}) 
\end{align}
In Fig.~\ref{fig:RFA-Renyi-Collection}, we illustrate the fractional entanglement signature of the MZM with respect to the second order Renyi entropy.
We see that all the main features of the fractional von Neumann entropy signature of the MZM compared to one of the bulk modes can be observed by the second order Renyi entropy.

The change of the quantum purity of the MZM saturates at a fractional value, but not exactly at $\log(2)/2$.
Instead, it saturates at a slightly larger value.
However, there is a clear difference between the saturation of the second order Renyi entropy for MZMs and bulk modes.
The second order R\'{e}nyi entropy of the bulk modes saturates at $\log(2)$, which is equivalent to the von Neumann entropy shown in the manuscript.\\
Performing the same analysis of the MZM and bulk modes using the RMI we obtain exactly the same results as for the von Neumann mutual information considered in the manuscript.
The MZM RMI is saturates at $\Delta I_2 = \log(2)$, while the bulk modes reach $\Delta I_2 = 2\log(2)$.

\end{document}